\documentclass[aps,prb,twocolumn,floatfix]{revtex4}

\usepackage{graphicx}
\usepackage{bm}
\usepackage{makeidx}
\usepackage{hyperref}
\def\bh{\bm H}
 
\begin{document}
\title{Piezomagnetic effect as a counterpart of negative thermal expansion in magnetically frustrated Mn-based antiperovskite nitrides}
\author{J.~Zemen,$^{1,2}$ Z.~Gercsi,$^{1,3}$ and K.G.~Sandeman$^{1,4,5}$}
\affiliation{$^{1}$Department of Physics, Blackett Laboratory, Imperial College London, London SW7 2AZ, United Kingdom}
\affiliation{$^{2}$School of Chemistry, University of Nottingham, University Park, Nottingham NG7 2RD, United Kingdom}
\affiliation{$^{3}$CRANN and School of Physics, Trinity College Dublin, Dublin 2, Ireland }
\affiliation{$^{4}$Department of Physics, Brooklyn College, CUNY, 2900 Bedford Ave., Brooklyn, NY 11210, USA }
\affiliation{$^{5}$The Graduate Center, CUNY, 365 Fifth Avenue, New York, New York 10016, USA }
\date{\today}

\begin{abstract}
Electric-field control of magnetization promises to substantially enhance the energy efficiency of device applications ranging from data storage to solid-state cooling. However, the intrinsic linear magnetoelectric effect is typically small in bulk materials. In thin films electric-field tuning of spin-orbit interaction phenomena (e.g., magnetocrystalline anisotropy) has been reported to achieve a partial control of the magnetic state. Here we explore the piezomagnetic effect (PME), driven by frustrated exchange interactions, which can induce a net magnetization in an antiferromagnet and reverse its direction via elastic strain generated piezoelectrically. Our $ab~initio$ study of PME in Mn-antiperovskite nitrides identified an extraordinarily large PME in Mn$_3$SnN available at room temperature. We explain the magnitude of PME based on features of the electronic structure and show an inverse-proportionality between the simulated zero-temperature PME and the negative thermal expansion at the magnetic (N\'eel) transition measured by Takenaka et al. in 9 antiferromagnetic Mn$_3$AN systems.
\end{abstract}

\maketitle

\section{Introduction}
\label{intro}
Emerging non-volatile magnetic random access memory (MRAM) devices represent bits of information as a magnetization direction which needs to be stabilised by magnetic anisotropy. A spin-transfer torque (STT) is typically used  to overcome the energy barrier between two stable directions. STT is induced by passing spin-polarized current which leads to Joule heating and sets limits on the storage density. Much research is focused on alternative switching mechanisms based on direct or indirect electric-field control of magnetic anisotropy which can reduce the dissipated energy by a factor of 100.\cite{matsukura2015control} 
Another recent alternative to STT-RAM devices replaces the ferromagnetic components with a single active antiferromagnetic (AFM) layer with a bistable alignment of the staggered moments. The switching then utilizes a spin-orbit torque (SOT) induced by an unpolarised electric current.\cite{wadley2016electrical,vzelezny2014relativistic} There is no dipolar coupling between neighbouring elements and they are insensitive to external magnetic fields. Again this alternative promises a higher storage density and energy efficiency. Note that both aforementioned alternatives to STT use the relativistic spin-orbit interaction (SOI) to achieve the thermal stability and the switching between distinct magnetic states. 

Here we explore an ambitious approach combining the electric-field control with the noncollinear antiferromagnetic structure of Mn-antiperovskite nitrides. 
The required coupling between the spin and orbital degrees of freedom is not due to the relativistic SOI but due to geometrically frustrated exchange interactions. The indirect magnetoelectric effect (ME) is hosted by a piezomagnetic Mn-antiperovskite layer elastically coupled to a piezoelectric substrate. We focus on the piezomagnetic effect (PME) which is characterised by a net magnetization directly proportional to the applied lattice strain.\cite{gomonaj1989magnetostriction,lukashev2008theory} Fully compensated AFM states are hard to track and utilize in general but the PME offers a valuable technique to probe and control the AFM ordering via the strain-induced magnetic moment.

In order to substantiate the future use of the PME in magnetoelectric composites, we perform a systematic $ab~initio$ study of PME in 9 cubic antiperovskites Mn$_3$AN (A = Rh, Pd, Ag, Co, Ni, Zn, Ga, In, Sn). We explain the variation of the magnitude of PME across this range of based on features of the electronic structure. The PME in Mn$_3$SnN predicted here is an order of magnitude larger than PME modelled so far in Mn$_3$GaN.\cite{lukashev2008theory} Moreover, the simulated PME is shown to be inversely proportional to the measured magnetovolume effect (MVE) at a magnetic (N\'eel) transition temperature\cite{takenaka2014magnetovolume} across the full set of 9 studied systems.
This agreement with experimental data is remarkable as both the PME and MVE originate in the frustrated AFM structure but we simulate PME at zero temperature whereas MVE was measured at the magnetic (N\'eel) transition temperature. MVE has not been modelled for this set of systems before. In addition to applications in spintronics our results can be used as a tool in search for materials with large negative thermal expansion (NTE) and barocaloric effect (BCE) which are both directly related to MVE.

Mn-based antiperovskite nitrides were first examined in 1970s.\cite{fruchart1971structure,fruchart1978magnetic} 
More recent experimental work on these metallic compounds includes a demonstration of large NTE in Mn$_3$AN (A = Ga, Zn, Cu, Ni)\cite{takenaka2005giant,wu2013magnetic,deng2015invar,deng2015frustrated} at the first order phase transition to a PM state. A large barocaloric effect was measured in Mn$_3$GaN at T$_N$ = 288~K\cite{matsunami2014giant} and the Mn-antiperovskites were consequently proposed as a new class of mechanocaloric materials. More importantly for spintronic applications, the baromagnetic effect (BME) closely related to the PME was reported in Mn$_3$G$_{0.95}$N$_{0.94}$ very recently,\cite{shi2016baromagnetic} the exchange bias effect was observed in Mn$_3$GaN/Co$_3$FeN bilayers,\cite{sakakibara2015magnetic} perpendicular magnetic anisotropy was demonstrated in Mn$_{67}$Ga$_{24}$N on MgO substrate, and the magnetocapacitance effect was measured in Mn$_3$GaN/SrTiO$_3$ bilayers.\cite{tashiro2013preparation}

Theoretical work on Mn-antiperovskites includes an early tight binding study\cite{jardin1981model} suggesting that the proximity of the Fermi energy to a sharp singularity (narrow N$_p$-Mn$_d$ band) in the electronic density of states has a large influence on the stability of the structural and
magnetic phases. However, this model considers only the nearest neighbour Mn-N hopping and neglects any hybridization with atom A. 
Phenomenological studies analysed phase transitions,\cite{gomonaj1992phenomenologic} magnetoelastic, and piezomagnetic\cite{gomonaj1989magnetostriction} properties with respect to the symmetry of the crystal and magnetic structure.
More recently $ab$~$initio$ modelling of the noncollinear magnetic structure has been carried out. The NTE and MVE are attributed to the frustrated exchange coupling between the three Mn atoms.\cite{qu2010nature,qu2012origin,deng2015invar}
The local spin density has been simulated for Mn$_3$GaN and Mn$_3$ZnN revealing its distinctly nonuniform distribution and localized character of the 3d~Mn moment.\cite{lukashev2010spin} The piezomagnetic\cite{lukashev2008theory} and flexomagnetic effect\cite{lukashev2010flexomagnetic} were simulated in Mn$_3$GaN by the same group. The strain-induced net magnetic moment predicted for Mn$_3$GaN is an order of magnitude lower than that of Mn$_3$SnN predicted in this work.

The PME is defined by a linear dependence of the net magnetization on elastic stress tensor components, in contrast to the magnetoelastic effect where the dependence on stress is quadratic. Both effects can be described phenomenologically by adding appropriate stress-dependent terms to the thermodynamic potential, i.e., the free energy:
\begin{equation} \label{Efree}
F(T, \bh, \sigma) = F_0(T, \bh) - \lambda_{i,jk} H_i \sigma_{jk} - \mu_{i,jk} H_i \sigma^2_{jk},
\end{equation}
where $\lambda_{i,jk}$ is an axial time-antisymmetric tensor representing the PME, $H_i$ are components of magnetic field, $\sigma_{jk}$ is the elastic stress tensor, and $\mu_{i,jk}$ is the magnetoelastic tensor. Non-vanishing elements of $\lambda_{i,jk}$  correspond to terms of eq.~(\ref{Efree}) which are invariant under operations from the magnetic symmetry group. \cite{borovik1994piezomagnetism} These elements then contribute to the magnetization:
 \begin{equation} \label{Mind}
M_i = -\frac{\partial F}{\partial H_i} = -\frac{\partial F_0}{\partial H_i} + \lambda_{i,jk} \sigma_{jk} + \mu_{i,jk} \sigma^2_{jk}.
\end{equation}

The PME was first proposed by Voigt\cite{voigt1928pme} in 1928. The linear character limits its existence to systems without time inversion symmetry or with magnetic group that contains time inversion only in combination with other elements of symmetry.\cite{tavger1956group} Hence, the PME is forbidden in all paramagnetic and diamagnetic materials. The most striking manifestation of PME is in antiferromagnets where the zero spontaneous magnetization acquires a finite value upon application of strain. The first AFM systems where PME was proposed\cite{dzialoshinskii1957problem,moriya1959piezomagnetism} and later observed\cite{borovik1959fluor} were transition-metal difluorides. In Mn-anitiperovskite nitrides PME was predicted quantitatively  in 2008\cite{lukashev2008theory} and it has not been observed experimentally so far.

\begin{figure}
\includegraphics[width=0.97\columnwidth]{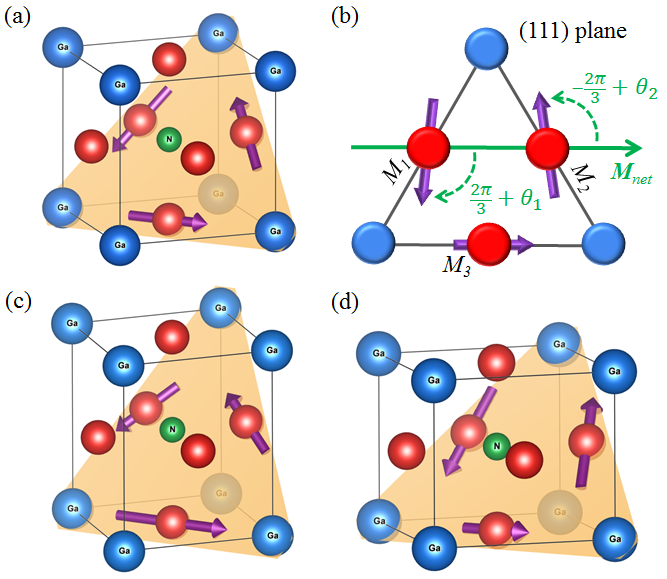}
\caption{(Color online) Mn-aniperovskite magnetic unit cell, cubic and strained lattice assuming Poisson's ratio of 0.5, the canting and changes of size are not to scale; (a)~unstrained structure of Mn$_3$GaN with local moments on Mn sites according to $\Gamma^{5g}$ representation; (b) tensile strained magnetic order in (111) plane, $M_{net}$ indicates the direction of the induced net moment; (c) compressively strained unit cell; (d) tensile strained unit cell.}
\label{f_structure}
\end{figure}

The noncollinear magnetic structure of Mn$_3$AN which hosts the PME and NTE considered in this work is shown in Fig.~\ref{f_structure}. (The direction of canting of the Mn local moments is specific for Mn$_3$GaN.) The ground state presented in Fig.~\ref{f_structure}(a) is the fully compensated AFM structure with symmetry corresponding to $\Gamma^{5g}$ representation.\citep{bertaut1968diffraction} (The magnetic unit cell belongs to the trigonal space group $P\overline{3}1m$ and has the same size as the cubic paramagnetic unit cell belonging to space group $Pm\overline{3}m$.) The exchange coupling between the neighbouring Mn atoms is antiferromagnetic which leads to the frustration in the triangular lattice in (111) plane (highlighted as orange online). The three Mn local magnetic moments (LMM) are of the same size and have an angle of $2\pi/3$ between their directions. A simultaneous rotation of all three LMMs by $\pi/2$ within the (111) plane results in another fully compensated AFM structure corresponding to $\Gamma^{4g}$ representation where the LMMs all point inside (outside) the triangle in a given (adjacent) plane.\citep{fruchart1971structure} The energy difference between $\Gamma^{4g}$ and $\Gamma^{5g}$ ordering is purely due to the spin-orbit coupling whereas the noncollinearity and magneto-structural coupling is due to the exchange interaction. It should be noted that the origin of PME in exchange interaction distinguishes it from magnetostriction which is due to spin-orbit coupling\cite{lukashev2008theory} (PME can be described as linear exchange-striction).  

\section{Results}
\label{results}

We calculate the total energy, magnetic moments, and projected density of states (DOS) for the noncollinear magnetic structure of biaxially strained Mn$_3$AN (A = Rh, Pd, Ag, Co, Ni, Zn, Ga, In, Sn) from first principles. Our computational procedure is the following:

(1) We find the equilibrium lattice parameter $a_0$, bulk modulus $K$, and the Poisson's ratio $\nu$ for each material with fixed AFM order by fitting the total energies obtained for a range of lattice parameters ($a, c/a$) to Birch-Murnaghan equation of state.\cite{birch1947finite} We also allowed for relaxation of individual atomic positions but we found no  
bond buckling in agreement with an earlier $ab~initio$ study.\cite{lukashev2008theory} The results are summarized in Table~\ref{tab}. 

\begin{figure}
\includegraphics[width=0.97\columnwidth]{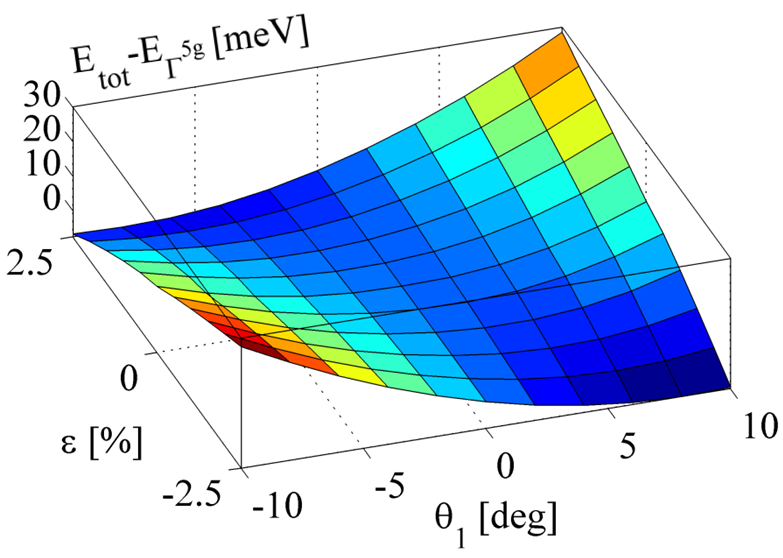}
\caption{(Color online) Total energy as a function of biaxial strain and canted angle for Mn$_3$GaN. No interpolation is used in the surface plot. The equilibrium angle depends linearly on the strain. The reference energy corresponds to $E(\theta_1=0)$ for each strain.}
\label{f_Etot}
\end{figure}

(2) We relax the magnetic moments with a fixed lattice for a range of biaxial strains to evaluate the PME. We perform two independent sets of calculations with the vertical lattice parameter $c$ set: (a) to conserve the unstrained unit cell volume - data labelled as~"V"; (b) according to the calculated Poisson's ratio - data labelled as~"P"; The initial AFM local moment directions and sizes are either relaxed by the VASP code\cite{kresse1999ultrasoft} in a self-consistent loop or explicitly by searching for minima in a total energy profile E$_{tot}(\epsilon,\theta_1)$ as shown in Fig.~\ref{f_Etot}. The quantitative agreement of these two methods gives us confidence that we found the physically relevant energy minimum. All calculations include the spin-orbit coupling and confirm that its impact on PME is negligible in case of period 4 and 5 elements.

(3) Finally, we increase the density of k-points and calculate the projected DOS for the converged strained and unstrained noncollinear structures in order to identify features in the electronic structure that would explain the variation of PME across the material range. Our results do not confirm  a proximity of the Fermi energy to a sharp peak in DOS as suggested by an earlier tight-binding study.\cite{jardin1981model}

Fig.~\ref{f_structure}(c) and (d) represent a qualitative overview of the simulated response of the magnetic structure to the tensile and compressive strain, respectively. A comparison with the ground state in Fig.~\ref{f_structure}(a) shows that Mn magnetic moments cant and change size which are two independent contributions to PME. This behaviour is due to the strain induced reduction of symmetry from $P\overline{3}1m$ to  $Pm'm'm$ orthorhombic magnetic space group and from $Pm\overline{3}m$ to $P4/mmm$ tetragonal space group in the paramagnetic case (the system is no longer invariant under the third order rotation about the (111) axis).  

For more clarity, Fig.~\ref{f_structure}(b) shows the tensile strained ($\epsilon = \Delta a/a_0 > 0$) magnetic order in the (111) plane. The canted angles $\theta_i$ within the (111) plane and LMM magnitudes M$_i$ on the three Mn sites are introduced. The moments in the (100) and (010) planes cant in opposite directions, $\theta_1 = - \theta_2$, to become more parallel (antiparallel) in case of positive (negative) $\theta_1$. The moment in the (001) plane does not change direction. 

The change of moment size $\Delta M_i = M_i-M_0$ is strongly dependent on the $c/a$ ratio of the tetragonal lattice. ($M_0$ is the LMM size common to all Mn sites in the unstrained system.) The changes plotted in Fig.~\ref{f_structure}(b) correspond to unit cell volume conservation when $\Delta M_1 = \Delta M_2 \approx - \Delta M_3/2$ for all studied systems. $M_3$ universally increases (decreases) with compressive (tensile) strain. With realistic Poisson's ratios all three Mn moments increase (decrease) for tensile (compressive) strain following the volume change of the unit cell (Mn$_3$RhN is the only exception where $M_3$ is almost independent on strain). Atom A develops a moment two orders of magnitude lower than the Mn local moment for small applied strain, $|\epsilon| < 1\%$, so its role in PME is negligible.

The unstrained ground state (plotted $\Gamma^{5g}$) has no spontaneous magnetization but a net moment $M_{net}$ aligned with M$_3$ develops upon straining. Our calculations confirm that the canted angle $\theta_i$, the change of moment size $\Delta M_i$, and consequently $M_{net} = 2 M_1 \cos(2\pi/3+\theta_1)+M_3$ depend linearly on applied strain as required by Eq.~(\ref{Mind}). The dependence departs slightly from linearity for larger strain $|\epsilon| > 1\%$, our study is limited to the interval $\epsilon \in \langle -2.5,2.5 \rangle \%$. A striking feature of PME is the change of orientation of $M_{net}$ when switching between tensile and compressive biaxial strain. Note that such control of net moment orientation cannot be achieved by magnetostriction. (The same description holds also for $\Gamma^{4g}$ order but $M_{net} \parallel M_3$ is then rotated by $\pi/2$ in (111) plane.)

\begin{table}[t]
\centering
\begin{tabular}{c|c c c c c c c} \hline
A  & $T_N$ [K]  & $a_0$ [\AA]  &  $a^t_0$ [\AA] & $\omega_s$ [$10^{-3}$] & $\nu^t$ & $K^t$ [GPa] & $M^t_0$ [$\mu_B$]   \\ \hline
Rh & 226 & 3.918 & 3.88 &  2.07 & 0.19 & 148.4 & 2.84 \\
Pd & 316 & 3.982 & 3.94 &  3.60 & 0.20 & 140.7 & 3.15 \\
Ag & 276 & 4.013 & 3.98 &  5.79 & 0.20 & 118.9 & 3.08 \\
Co & 252 & 3.867 & 3.80 &  5.64 & 0.13 & 149.5 & 2.48 \\
Ni & 256 & 3.886 & 3.84 &  8.18 & 0.15 & 136.5 & 2.83 \\
Zn & 170 & 3.890 & 3.87 & 20.44 & 0.13 & 126.0 & 2.64 \\
Ga & 288 & 3.898 & 3.86 & 19.10 & 0.13 & 129.4 & 2.43 \\
In & 366 & 4.000 & 3.99 &  9.24 & 0.18 & 115.0 & 2.70 \\
Sn & 475 & 4.060 & 3.97 &   0.0 & 0.18 & 102.0 & 2.52 \\ \hline
\end{tabular}
\caption{Physical properties of Mn$_3$AN: N\'eel temperature, lattice parameter at 10~K, calculated lattice parameter, spontaneous volume change, Poisson's ratio, bulk modulus, size of Mn local moment in unstrained system; all measured data are taken from Ref.~[\onlinecite{takenaka2014magnetovolume}] except $a_0$ and $T_N$ for Mn$_3$SnN which are from Ref.~[\onlinecite{LB1981}]. Calculated data are marked$^t$.}
\label{tab}
\end{table}

Table~\ref{tab} list all relevant measured properties and results calculated in this work. Our Mn magnetic moment for Mn$_3$GaN is in good agreement with a previous theoretical study.\cite{lukashev2008theory} 
Our Poisson's ratios do not vary much across the range of compounds and are slightly smaller than $\nu$=0.25-0.3 predicted by an $ab~initio$ study of elastic properties in Mn$_3$(Cu,Ge)N.\cite{qu2011elastic} All calculated lattice parameters are 1-2\% smaller than the values measured at low temperatures.

Fig.~\ref{f_PME} presents our results on PME and the related features of electronic band structure. The net moment $M_{net}$ plotted for the nine Mn-antiperovskite systems subject to tensile strain $\epsilon = 1\%$ is a natural measure of PME. Positive (negative) value of $M_{net}$ corresponds to net moment induced parallel (antiparallel) to $M_3$ irrespective of belonging to the $\Gamma^{4g}$ or $\Gamma^{5g}$ representation. 

Fig.~\ref{f_PME}(a) compares the PME obtained assuming unit cell volume conservation (Poisson's ratio $\nu = 0.5$) and using our calculated Poisson's ratios, $\nu$, listed in Table~\ref{tab} which correspond to smaller vertical distortion for a given strain. The latter is our lower estimate of the experimentally accessible PME as our calculated values of $\nu$ are lower than expected for metallic materials. The former version of PME neglects the elastic properties of the lattice and represents the response of the frustrated magnetic system to a lattice symmetry breaking (normalized tetragonal distortion). As a result, the predicted $M_{net}$~(V) should be regarded as an upper estimate of the experimentally accessible PME. In both cases Mn$_3$SnN is predicted to have $M_{net}$ an order of magnitude larger than Mn$_3$GaN, the only PME value available in literature.\cite{lukashev2008theory} 

\subsection{Fitting PME by Heisenberg model}

In order to interpret the calculated PME in terms of the AFM pairwise exchange interactions $J_{ij}(\epsilon)$ between the three Mn atoms in the (111) plane we resort to the classical Heisenberg model:
\begin{eqnarray} \label{Heisenberg}
E(\theta_1,\epsilon) & = & -J_{12}M_1M_2\cos(2\pi/3-2\theta_1) \nonumber \\ 
            &   & -2J_{13}M_1M_3\cos(2\pi/3+\theta_1), \;
\end{eqnarray}
where the values of the exchange parameters $J_{13} = J_{23} \neq J_{12}$ and the local moments $M_1 = M_2 \neq M_3$ introduced in Fig.~\ref{f_structure}(b) are restricted by the tetragonal symmetry. We find the canted angle minimizing the exchange energy ($\partial E / \partial \theta_1 = 0$) and insert it into the expression for the net moment $M_{net} = 2 M_1 \cos(2\pi/3+\theta_1)+M_3$. We obtain a relationship between PME and changes of the exchange interaction due to strain:
\begin{eqnarray} 
\frac{M_{net}}{M_3} & = & 1 - \frac{J_{13}}{J_{12}} \label{J13J12} \\
 & \approx & \frac{J_0 - \Delta J -(J_0 + \Delta J)}{J_0 - \Delta J} \approx -\frac{2\Delta J}{J_0} \nonumber \\
M^J_{net} & \equiv & -\frac{2M_3}{J_0} \Delta J = \frac{2M_3}{J_0}\frac{\partial J_{12}}{\partial \epsilon} \Delta \epsilon, \label{dJde} \;
\end{eqnarray}
where $J_0<0$ is the exchange parameter in the unstrained lattice and $\Delta J$ is the induced change of $J_{12}$ and $J_{13}$. We fitted our $ab~initio$ total energy as a function of the canted angle to the Heisenberg model of eq.~(\ref{Heisenberg}) to extract $J_{12}$ and $J_{13}$ for each value of strain. In all compounds we observed: $J_{12} \approx J_0 - \Delta J$ and $J_{13} \approx J_0 + \Delta J$ which allows us to define $M^J_{net}$ in eq.~(\ref{dJde}) that is directly proportional to the derivative of the exchange parameters $J_{ij}$ with respect to the biaxial strain $\epsilon$. 

\begin{figure}
\includegraphics[width=0.97\columnwidth]{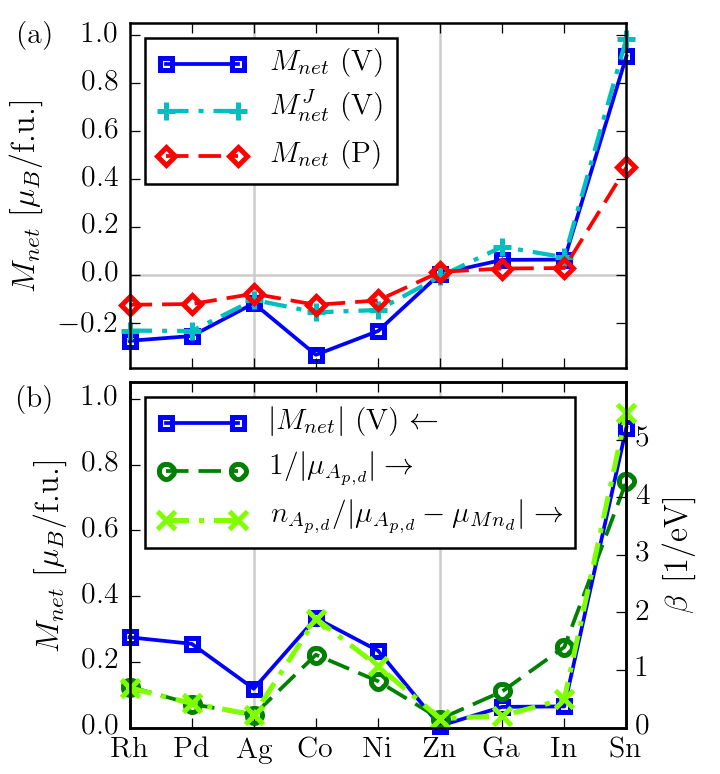}
\caption{(Color online) Comparison of the net moment $M_{net}$ induced by 1\% of tensile strain: (a) $M_{net}$ assuming unit cell volume conservation (V) and Poisson's rations of Table~\ref{tab} (P) and $M^J_{net}$ fitted according to eq.~(\ref{dJde}); (b) comparison of PME, measured by $M_{net}$(V), to the inverse of the energy separation between $p$ or $d$-states of atom A and $d$-states of Mn (weighted by the relative band filling), marked as $\beta$.}
\label{f_PME}
\end{figure}

Fig.~\ref{f_PME}(a) shows that $M^J_{net}$ is in good agreement with $M_{net}$~(V) extracted directly from our calculated LMMs (without any fitting). The small differences are due to deviations of the magnetic system from the Heisenberg behaviour (e.g., LMMs change size as they cant even in an unstrained lattice) and deviations from linearity assumed in eq.~(\ref{J13J12}). The key conclusion based on Fig.~\ref{f_PME}(a) in combinations with eq.~(\ref{J13J12}) is that a large PME corresponds to a large difference between $J_{12}$ (bond in the plane of the biaxial strain) and $J_{13}$ (bond with a component perpendicular to this plane.)

\subsection{Linking PME to band structure}

Fig.~\ref{f_PME}(b) relates the total induced moment $M_{net}$~(V) to the mean band energy of the valence $p$ or $d$-states of atom A. This quantity is often called the band center and we extract it from our projected DOS, $\rho_{A_{p,d}}(E)$, as follows: $\mu_{A_{p,d}} = 1/\Omega \int E\rho_{A_{p,d}}(E)dE$, where $\Omega = \int \rho_{A_{p,d}}(E)dE$ is a normalization. We consider only the $d$-band ($\rho_{A_d}(E)$) when atom A is a transition metal and only the $p$-band ($\rho_{A_p}(E)$) for the rest. The wide $s$-band does not seem to play an important role in PME. The right vertical axis of Fig.~\ref{f_PME}(b) measures the inverse of $\mu_{A_{p,d}}$ with respect to the Fermi energy ($E_F$), marked as $\beta_1 \equiv 1/|\mu_{A_{p,d}}|$, and the same quantity with respect to the Mn $d$-band center, $\mu_{Mn_d}$, weighted by the relative occupation of $p$ or $d$-band of atom A, marked as $\beta_2 \equiv n_{A_{p,d}}/|\mu_{A_{p,d}}-\mu_{Mn_d}|$, where $n_{A_{p,d}} = 1/N\int^{E_F} \rho_{A_{p,d}}(E)dE$ and $N$ is the occupation of a fully filled $p$ or $d$-band. 

Based on the remarkable match between $|M_{net}|$ and both variants of $\beta_{1,2}$ we conclude that piezomagnetism in Mn-antiperovskite nitrides is governed by the mutual configuration of Mn $d$-states and $p$ or $d$-states of atom A. More specifically, a greater proximity (a potential for hybridization) of the valence band of atom A to the spin-polarized $d$-band of Mn increases the difference between $J_{12}$ and $J_{13}$ per unit strain which manifests itself as a larger induced net moment. On the other hand, when the triangular magnetic order of Mn moments is undisturbed by hybridization with $p$ or $d$-states of atom A then 
$J_{12} \approx J_{13}$ and only a small net moment is induced. The best example is Mn$_3$ZnN where the narrow fully filled $d$-band is about 7~eV below the Fermi energy and the induced net moment is negligible. 
This trend is analogous to a scaling of the N\'eel temperature with the number of valence electrons of atom A in the same class of materials detected in 1977.\cite{fruchart1978magnetic}

It should be noted that Mn$_3$AgN and Mn$_3$RhN do not share the triangular AFM order according to earlier neutron diffraction studies,\citep{LB1981} whereas the magnetic structure of Mn$_3$CoN and Mn$_3$PdN is yet to be confirmed experimentally. We include these four compounds in our study as their composition, AFM order, and experimentally resolved MVE\citep{takenaka2014magnetovolume} makes them potential candidates for piezomagnetic behaviour. 

In more general terms, we perform a computational experiment when the magnetic system is initialized in the triangular state ($\Gamma^{4g}$ or $\Gamma^{5g}$) even if it was only a local energy minimum for Mn$_3$AN (A = Ag, Co, Pd, Rh) and the response (induced $M_{net}$) to a tetragonal distortion is detected. The consistency of the piezomagnetic response across the whole set of materials motivates us to use this procedure as a probe of the level of frustration of the exchange interaction even if the real systems did not host piezomagnetism.
In the following paragraphs, we compare our simulated PME to the spontaneous magnetovolume effect which is a measure of the magnetic frustration and experimental data is available for all nine Mn$_3$AN compounds.\cite{takenaka2014magnetovolume}

\subsection{Comparing PME to MVE}

To draw an analogy between the strain and an external field $H$ that can induce magnetization, we introduce a piezomagnetic susceptibility:
\begin{equation} \label{suscep}
\frac{M^J_{net}}{M_3} = \frac{2}{J_0} \frac{\partial J_{12}}{\partial \epsilon} \Delta \epsilon \equiv \chi_P(\mu_{A_{p,d}}) \Delta \epsilon,
\end{equation} 
where the change of applied train $\Delta \epsilon$ replaces $H$ and   $M^J_{net}(\Delta \epsilon)$ was introduced in eq.~(\ref{dJde}). Based on Fig.~\ref{f_PME}(b) we can say that the susceptibility $\chi_P(\mu_{A_{p,d}})$ is inversely proportional to the mean valence band energy of atom A in the unstrained system. 

\begin{figure}
\includegraphics[width=0.97\columnwidth]{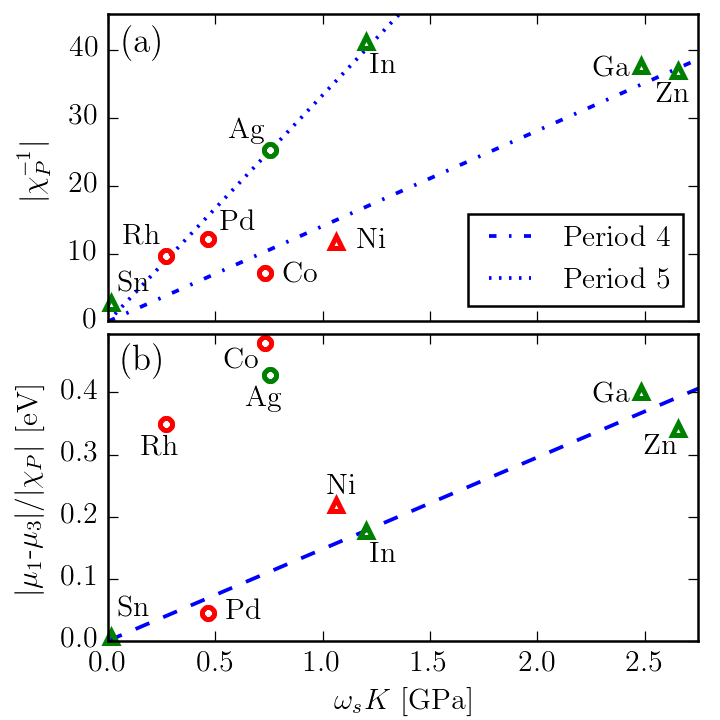}
\caption{(Color online) Calcualted PME characterised by $|\chi_P^{-1}|$ as a function of the measured MVE weighted by bulk modulus, triangles indicate systems with confirmed triangular magnetic ground state, red and green symbols indicate a positive and negative canted angle at tensile strain, respectively; the blue lines are least square linear fits; (a)~two different trends for atom A from period 4 and 5; (b)~$|\chi_P^{-1}|$ weighted by a strain induced shift of mean band energy of two Mn atoms - one trend for all systems with confirmed triangular magnetism.}
\label{f_MVE}
\end{figure}

Fig.~\ref{f_MVE} compares the measured magnetovolume effect\citep{takenaka2014magnetovolume} to our calculated piezomagnetic susceptibility $\chi_P$. 
The MVE is a spontaneous change of volume due to a change of magnetic ordering (typically the size of magnetic moment). It was first observed in Ni-Fe Invar below its $T_C$.\cite{hayase1973spontaneous} Takenaka et al. measure a spontaneous volume increase upon the transition from PM to AFM state and subtract the phononic contribution so their MVE data are purely of magnetic origin.\citep{takenaka2014magnetovolume} They investigate a wide range of Mn-antiperovskite nitrides and conclude that MVE is a property of the frustrated triangular AFM state which is strongly dependent on the number of valence electrons. MVE is the largest when there are two $s$-electrons and one or no $p$-electrons (A = Zn, Ga). When the number of valence $s$ and $p$-electrons changes then the systems transforms to a different crystal/magnetic structure with no MVE (A = Cu, Ge, As, Sn, Sb). 

In addition, Takenaka et al. have observed an increase in MVE as the $d$-band of atom A moves away from $E_F$. This general trend reminds us of the scaling of susceptibility $\chi_P$ with the mean band energy of atom A $\mu_{A_{p,d}}$ described above. We include Fig.~\ref{f_MVE}(a) to check if the dependence on $\mu_{A_{p,d}}$ furnishes a clear link between PME and MVE. The figure shows that our piezomagnetic susceptibility $\chi_P$ is inversely proportional to the measured volume change as expected. In other words, a large MVE implies a small PME and vice versa. Atoms A belonging to periods 4 and 5 of the periodic table have different coefficients of proportionality. This implies that not only the position of A-band with respect to $E_F$ but also the size of atom A plays a role in weakening the triangular AFM structure. Such difference between period 4 and 5 was first seen also in case of the scaling of $T_N$ with the number of  valence electrons of atom A in 1977.\cite{fruchart1978magnetic}

The agreement of a calculated zero temperature susceptibility ($\chi_P^{-1}$) with a spontaneous volume change $\omega_s$ at the PM-AFM phase transition (weighted by $K$) is remarkable and requires further analysis. Magnetovolume effects in itinerant electron magnets were first analysed by the Stoner-Edwards-Wohlfarth theory.\citep{wohlfarth1977thermodynamic} The free energy can be approximated by $F(T,M,\omega) = F_0(T,\omega) + \frac{1}{2}KV\omega^2 + \frac{1}{2}a(T,\omega)M^2$ and minimized with respect to the volume strain $\omega=\Delta V/V$ to obtain $KV\omega = c_{mv}M^2$ where $c_{mv} = -\frac{1}{2}\partial a(T,\omega)/\partial \omega$ is the magnetovolume coefficient, $M$ is the spontaneous magnetization, $K$ is the bulk modulus, and $V$ is the reference volume. After considering the spin fluctuations at the first-order phase transition the above formula becomes: $KV\omega = c_{mv}(M^2-\xi^2)$ where $\xi$ is the amplitude of spin fluctuations.\cite{moriya1980magneto,takahashi2006magnetovolume}

In the case of Mn-antiperovskites 
the local moments are relatively well localized\cite{lukashev2010spin} so we can approximate the magnetic energy of the triangular AFM system on a cubic lattice by eq.~(\ref{Heisenberg}) with zero canted angle: $E(\theta_1=0) = \frac{3}{2}J_0M_0^2$. The balance of elastic and magnetic energy then leads to an expression for the spontaneous volume strain ($\Delta V/V$):
\begin{equation} \label{tv}
\omega_s K = -\frac{3M_0^2}{2V}\frac{\partial J_0}{\partial \omega} \sim \frac{\partial J_0}{\partial \omega} \equiv t^v,
\end{equation}
where we neglect the change of local moments $M_0$ with changing volume, $(\Delta M_0)^2$, as a higher order contribution. 
The magnetic stress per Mn-Mn bond $t^v$ is introduced following the work of Filippetti and Hill.\cite{filippetti2000magnetic} The magnetic stress at the phase transition can be then expressed as: $T^v = \partial E^{\Gamma^{5g}}/\partial \omega = \frac{3}{2}t_vM_0^2$, where $E^{\Gamma^{5g}}$ is again the magnetic energy $E(\theta_1=0)$. 

After establishing the link between MVE and the magnetic volume stress $T^v$, we attempt the same for PME and the magnetic biaxial stress:
$T^b = \partial E(\theta_1)/\partial \epsilon \sim t^b$, where $E(\theta_1)$ is a magnetic energy of the canted AFM structures and the magnetic stress per Mn-Mn bond $t^b$ is proportional to the susceptibility $\chi_P$ of eq.~(\ref{suscep}):
\begin{equation} \label{tb}
\chi_P = \frac{2}{J_0} \frac{\partial J_{12}}{\partial \epsilon} \sim \frac{\partial J_{12}}{\partial \epsilon} \equiv t^b.
\end{equation}

Finally, based on the comparison of eqs.~(\ref{tv}) and (\ref{tb}) we can conclude that both $\omega_s K$ and $\chi_P$ are proportional to derivatives of the exchange parameters with respect to strain and thereby to the magnetic stress of the triangular AFM system. Hence the linear relationship of Fig.~\ref{f_MVE}(a) indicates a trade-off between two complementary stress relief mechanisms.

\section{Discussion}
\label{se_discussion}

In principle, the stress arising at the onset of AFM ordering at $T_N$ can be relieved by a volume change or a lattice distortion. However, our calculations and subsequent fitting to Heisenberg model find that the magnetic energy saved by a tetragonal distortion (linear in $\epsilon$) becomes smaller than the elastic energy cost (quadratic in $\epsilon$ around unstrained lattice) for negligibly small distortions. This is confirmed by x-ray diffraction\citep{takenaka2014magnetovolume} which has not indicated a tetragonal distortion in any compound studied in this work. Nevertheless, $\chi_P$ reflects how much magnetic stress could be relieved by a tetragonal distortion and this quantity is inversely proportional to $\omega_s K$ as shown by Fig.~\ref{f_MVE}. We plot $\chi_P$ vs $\omega_s K$ rather than $\omega_s$ to compare only quantities related to magnetism and factor out the system dependent elastic properties.

It should be noted that the sign of $\chi_P$ indicates which type of tetragonal distortion is energetically more favourable. A brief demonstration of this neglects the dependence of $M_i$ and $\theta_i$ on strain in eq.~(\ref{Heisenberg}) - then we can find a spontaneous biaxial strain $\epsilon_s$ (analogous to volume strain $\omega_s$) from the balance of elastic and magnetic energy: $\epsilon_s = \partial J_{12}/\partial \epsilon M_0^2/C = -\frac{1}{2}\chi_P|J_0|M_0^2/C$, where $C>0$ is an effective elastic modulus. Immediately, we can see that all systems in this study with $\chi_P>0$ tend to a distortion with $\epsilon_s<0$ ($c/a>1$) and vice versa.

We conclude that a system with robust triangular magnetic order undisturbed by the proximity of electronic states of atom A (large $\mu_{A_{p,d}}$) tends to relieve its magnetic stress via a volume change, whereas a system more influenced by atom A but with persisting triangular order (small $\mu_{A_{p,d}}$) prefers to relieve its magnetic stress via a tetragonal distortion should the elastic energy cost allow it. (If the tetragonal distortion is enforced externally, then the system develops a large net magnetization.) 

The slight deviations of $|\chi_P^{-1}|$ from $\omega_s K$ seen in Fig.~\ref{f_MVE} may originate in: (a) spin fluctuations which we neglected in eq.~(\ref{tv}), the small size of the deviations suggests that the spin fluctuation contribution to MVE ($KV\omega = c_{mv}(M^2-\xi^2)$) is significantly suppressed by the strong frustration; (b) limited numerical accuracy, e.g., Mn$_3$ZnN is most affected as it has almost trivial $\chi_P$ and its large relative error is amplified by the inversion; (c) Nitrogen deficiency (8-16\%) varying across the range of samples where MVE was measured,\citep{takenaka2014magnetovolume} e.g., magnetic order in Mn$_3$SnN is known to be sensitive to N concentration; \citep{LB1981} (d) a material-specific elastic property that was not factored out of the plotted quantities, e.g., the use of bulk modulus $K=130$~GPa for all compounds when subtracting the phononic contribution to MVE\citep{takenaka2014magnetovolume} (consequently, in the plot we use $K=130$~GPa instead of our calculated $K$ of Table~\ref{tab}).

To further explore the inverse proportionality between PME and MVE with respect to features of the electronic structure we analyse the strain dependence of mean band energy of Mn-states. We extract the mutual shift of mean band energy of Mn$^1_d$-states (site in (100) plane of the unit cell) and Mn$^3_d$ (site in (001) plane) from the projected DOS $\rho_{Mn^1_d}(E,\epsilon,\theta_1)$ and $\rho_{Mn^3_d}(E,\epsilon,\theta_1)$ of the strained system before canting($\epsilon$=1\%, $\theta_1$=0) in analogy to evaluation of $\mu_{A_{p,d}}$ shown in Fig.~\ref{f_PME}(b). The obtained quantity $|\mu_{1}-\mu_{3}|$ directly measures the response of the spin polarized electronic structure to the tetragonal distortion. Such information is missing in $\mu_{A_{p,d}}$ of the unstrained structure.

Fig.~\ref{f_MVE}(b) shows $|\chi_P^{-1}|$ weighted by the mutual band shift $|\mu_{1}-\mu_{3}|$ as a function of $\omega_s K$. Compounds with atom A from period 4 and 5 now follow the same linear trend with the exception of A = Ag, Co, Rh. Our hypothesis based on Fig.~\ref{f_MVE} is that the factor $|\mu_{1}-\mu_{3}|$ incorporates the dependence of PME on the size of atom A for systems with stable triangular AFM ordering. Mn$_3$AgN and Mn$_3$RhN do not have triangular AFM ground state which has explanation in their band structure properties and become apparent in Fig.~\ref{f_MVE}(b). Extending the same argument to the unknown magnetic order, we expect Mn$_3$PdN (Mn$_3$CoN) to have a triangular (other) AFM ground state.

The linear scaling of the spontaneous MVE with $|\chi_P^{-1}|$ implies a significant suppression of spin fluctuations by the strong frustration in these systems. At the same time it can be used as a tool in theory led design of non-stoichiometric materials with large MVE and consequently BCE where the entropy change is proportional to the spontaneous volume change according to the Clausius-Clapeyron relation:
\begin{equation} \label{entropy}
S(T_t,p)-S(T_t,0) = V \omega_s \left( \frac{dT_t}{dp} \right)^{-1}.
\end{equation}
Modelling the pressure dependence of the transition temperature $dT_t/dp$ goes beyond the capability of density functional theory at zero temperature and is the subject of our ongoing work.\cite{zemen2016frustrated}
 
We hope that the successful comparison of our predicted PME to the measured MVE and the coherent interpretation of the PME based on features of the electronic structure will provide guidance for further investigations of the unique physical properties of the frustrated AFM structure of Mn-antiperovskites and enable development of applications including data storage, memory, and solid-state cooling.

\section{Methods}
\label{se_methods}

All our calculations employ the projector augmented-wave (PAW) method
implemented in VASP code\cite{kresse1999ultrasoft} within the Perdew-
Burke-Ernzerhof (PBE) generalized gradient
approximation.\citep{perdew1996generalized} 
This approach allows for relaxation of fully unconstrained noncollinear magnetic structures.\citep{hobbs2000fully} We use a 12x12x12 k-point sampling in the self-consistent cycle and 17x17x17 k-point sampling to obtain the site and orbital resolved DOS. The cutoff energy is 400~eV. The local magnetic moments are evaluated in atomic spheres with the default Wigner Seitz radius as they are not very sensitive to the projection sphere radius.\cite{lukashev2008theory}

We constrain the Mn local moment directions using an additional penalty energy as implemented in the VASP code in order to obtain the projected DOS $\rho_{Mn^3_d}(E,\epsilon,\theta_1)$ of the strained system. We add a further constraint to suppress the small moment on atom A which develops due to strain to allow for extraction of $J_{12}$ and $J_{13}$ from the total energy as a function of strain and canted angle by fitting to the Heisenberg model of eq.~(\ref{Heisenberg}).
\acknowledgments

We would like to thank Kirill Belashchenko, Lesley Cohen, and Julie Staunton for productive discussions.
The research leading to these results has received funding from the European Community’s 7th Framework Programme under Grant agreement 310748 “DRREAM”.


\bibliography{magnetocaloric}
\end{document}